\documentclass[prl,twocolumn,showpacs,preprintnumbers,amsmath,amssymb]{revtex4}

\usepackage{graphicx}
\usepackage{dcolumn}
\usepackage{bm}

\begin{document}

\preprint{PRL/Oest} \draft
\title{Spin noise spectroscopy in GaAs}
\author{M. Oestreich}
\altaffiliation{M.O. is currently at: UCSB, Materials Department,
Santa Barbara CA93103-5050, USA} \affiliation{Institut f\"ur
Festk\"orperphysik, Universit\"at Hannover, Appelstra\ss e 2,
D-30167 Hannover (Germany)}

\author{M. R\"omer }
\affiliation{Institut f\"ur Festk\"orperphysik, Universit\"at
Hannover, Appelstra\ss e 2, D-30167 Hannover (Germany)}

\author{R. J. Haug}
\affiliation{Institut f\"ur Festk\"orperphysik, Universit\"at
Hannover, Appelstra\ss e 2, D-30167 Hannover (Germany)}

\author{D. H\"agele}
\affiliation{Institut f\"ur Festk\"orperphysik, Universit\"at
Hannover, Appelstra\ss e 2, D-30167 Hannover (Germany)}

\date{\today}

\begin{abstract}
We observe the noise spectrum of electron spins in bulk GaAs by
Faraday rotation noise spectroscopy. The experimental technique
enables the undisturbed measurement of the electron spin dynamics
in semiconductors. We measure exemplarily the electron spin
relaxation time and the electron Land\'{e} g--factor in $n$--doped
GaAs at low temperatures and find good agreement of the measured
noise spectrum with an unpretentious theory based on
Poisson distribution probability.\\
\end{abstract}


\maketitle

The inexorable decrease of structure size in semiconductor devices
inevitably leads from today's quasi-classical devices to quantum
mechanical devices. These quantum mechanical devices might rely
not only on the charge of electrons, i.e. on the spatial part of
the electron wave function, but also on the much more robust spin
part of the wave function. The robustness of the electron spin
motivates the current extensive research on the spin dynamics in
semiconductors and might lead to spintronic devices with superior
functionality and to the enchaining goal of spin quantum
information processing
\cite{science294_1488,nature430_431,dyakonov2002}.

One important signature of the spin dynamics in semiconductors
results from the thermal fluctuations of electron spin occupation
in the conduction band which fluctuates on the time scale of the
spin lifetime and gives rise to spin noise. This kind of spin
noise has been observed recently in rubidium gas atoms
\cite{nature431_49} and theoretically exploited for spin currents
through single quantum dots \cite{prl94_06601}. The experimental
observation of spin noise in semiconductors is however a major
challenge since the photon shot noise in optical experiments and
the Coulomb shot noise in electrical experiments is usually orders
of magnitude larger than the spin noise. On the other hand,
measurement of the spin noise in semiconductors has the power to
become an important experimental tool since the spin noise
spectrum yields not only information about the electron Land\'{e}
g--factor but also promises to give insight into
electron--electron spin correlations, spin phase transitions, and
spin fluctuations in low--dimensional semiconductor structures.
Additionally, spin noise spectroscopy has the advantage to detect
optically the spin dynamics in semiconductors without optical
excitation of electrons and holes.

\begin{figure}
\includegraphics[width=8cm]{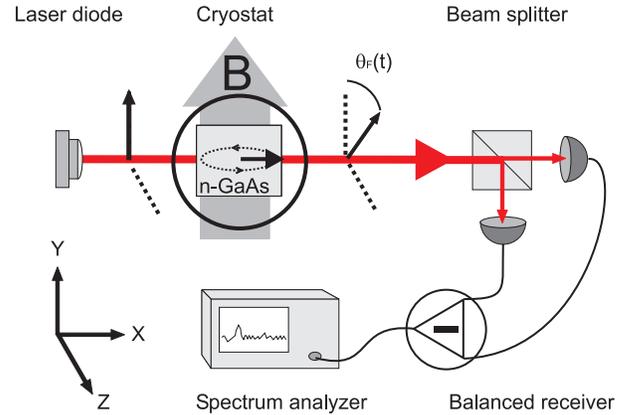}
    \caption{ \label{fig0}(color online) Schematic of the experiment.
    Thermodynamic fluctuations of the electron spins in $n$--doped
    GaAs precess around an external magnetic field. The precessing
    spin fluctuations cause oscillations of the Faraday rotation
    signal $\Theta_{\rm F}$
    whose power spectrum is detected by a
    balanced receiver and a spectrum analyzer.
    }
\end{figure}

This Letter presents an experimental and theoretical description
of the spin noise in semiconductors. The experimental setup is
schematically depicted in Fig.~\ref{fig0}. The optical
measurements are performed on a 370~$\mu$m thick GaAs wafer with a
silicon $n$--doping of $1.8\times 10^{16}$~cm$^{-3}$ by
Faraday--rotation spectroscopy. The light of a cw laser diode is
sent through a spatial filter, linearly polarized by a Glan
Thompson polarizer, and focused  on the GaAs sample which is
mounted in Voigt geometry in a superconducting split--coil magnet
with variable temperature insert. The laser wavelength is tuned
10~nm below the GaAs band-gap to minimize absorption and maximize
the Faraday rotation signal. The focus diameter of the laser is
about 65~$\mu$m. The linearly polarized light passing through the
sample is split by a polarizing beam splitter into two components,
linearly polarized $\pm 45^\circ$ with respect to the initial
polarization. The two components are focused on a pair of photo
diodes of a NewFocus 650~MHz balanced photo receiver and the time
varying difference of the two - equally strong if temporally
averaged
 - components is converted with a gain of 350 V/W into voltage
and measured by an HP spectrum analyzer. The laser intensity
before the beam splitter is 1.9~mW resulting in a white photon
shot noise of $10$~${\rm nV}/\sqrt{\rm Hz}$. The electrical noise
of the combination of balanced receiver and spectrum analyzer is
$13$~${\rm nV}/\sqrt{\rm Hz}$ \cite{note_electrical_noise}. We can
distinguish between the large external noise sources and the small
spin noise by applying a weak magnetic field to the sample
perpendicular to the direction of the laser light propagation. The
spin noise strongly depends on the magnetic field since any
statistical imbalance of the electron spin ensemble precesses
around the magnetic field resulting in a spin noise peak at the
precession frequency. The width of the spin noise peak is
proportional to the spin relaxation rate, which includes diffusion
of the thermal spin fluctuations out of the laser focus. The white
photon and electron noise powers do not depend on the magnetic
field and therefore can be easily subtracted.

\begin{figure}
\includegraphics[width=8cm]{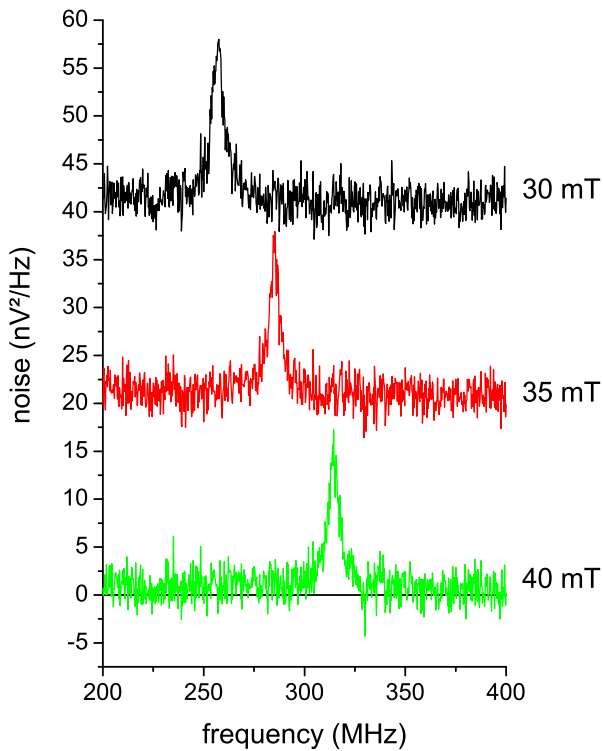}
    \caption{ \label{fig1} (color online) Spectrum of the spin noise
    for different applied magnetic fields. The white photon shot noise and
             electrical noise power are subtracted. The spectra are vertically
             shifted for clarity. The sample temperature is 10~K.
             Frequency and width of the spin
             noise spectrum are a direct measure of the electron
             Land\'{e} g--factor and the electron spin relaxation time,
             respectively.}
\end{figure}

Figure~\ref{fig1} depicts the measured spin noise spectra for
three magnetic fields. The power spectrum at each magnetic field
is averaged over 62 measurements \'a 10 minutes and subtracted by
interleaving 62 measurements \'a 10 minutes at zero magnetic field
to subtract the photon and electrical noise. We alternate between
finite and zero magnetic field to eliminate any influence of
thermal drift in the electronics. The spin noise maxima in
Fig.~\ref{fig1} shift linearly with $B$ since the spin precession
frequency and therefore the spin noise frequency is directly
proportional to $B$. The width of the spin noise spectra is about
7~MHz yielding a spin relaxation time of about 45~ns. This spin
relaxation time is consistent with earlier measurements by Dzhioev
et al. who measure 50~ns for localized donor bound electrons in
GaAs with an $n$--doping of $1.5\times 10^{16}$~cm$^{-3}$
\cite{prb66_245204}.

\begin{figure}
\includegraphics[width=8cm]{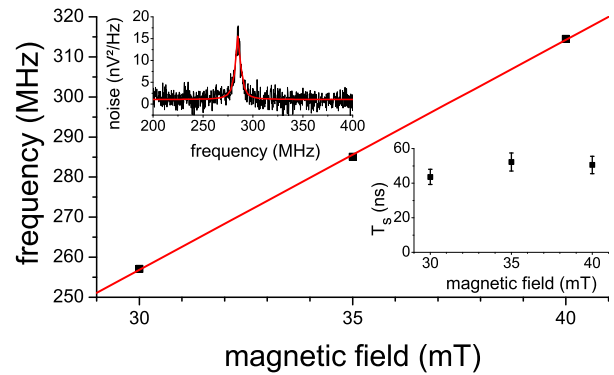}
    \caption{ \label{fig2} (color online) Frequency of the maximum of the spin
    noise (filled squares) and spin relaxation time $T_s$ determined
    from the full--width half--maximum (right inset) versus magnetic
    field $B$. The slope of the linear fit to the frequency (solid line) yields
    an electron Land\'{e} g--factor of $-0.41\pm 0.01$. The fit does not extrapolate to
    $f\rightarrow 0$ for $B\rightarrow 0$ due to a known constant remanent
    field of the superconducting magnet, which was not subtracted in this figure.
    The upper, left inset depicts the Lorentzian fit to the noise spectra at 35~mT.}
\end{figure}

Figure~\ref{fig2} depicts the maxima of the spin noise peaks
versus $B$. The spin precession frequency is equal to $g_e \mu_B
B/h$, where $g_e$ is the electron Land\'{e} g--factor, $\mu_B$ is
Bohr's magneton, and $h$ is Planck's constant. The  slope of the
measured noise frequency $f$ versus $B$ yields $g_e=-0.41\pm 0.01$
which is in excellent agreement with optically oriented and
detected electron spin resonance in lightly doped $n$--GaAs
\cite{prb67_165315} confirming that the measured spin noise is in
fact correlated with donor bound electrons. The width of the noise
spectra is a measure of the spin relaxation time (right inset in
Fig.~\ref{fig2}), which is within the measurement error
independent of $B$, i.e., the spin relaxation measurement is not
obscured by inhomogeneous broadening of $g_e$.

In the following, we want to theoretically estimate the amplitude
of the spin noise and compare the results with our measurements.
We make the simplifying assumption that the statistical
fluctuation of the electron spin polarization along the light
propagation is proportional to $\sqrt{N}$, where $N$ is the number
of donor bound electrons in the volume $V$ of the laser. In our
experiment, this is a good approximation since the electrons are
localized at the donors and the Poisson prerequisite of
independent events is therefore fulfilled. The experiment
fortifies the assumption since we do not observe a temperature
dependence of the width and the height of the noise peak when we
increase the temperature from 5~K to 10~K, i.e. the area under the
noise peak stays constant. The localization of the electrons has
additionally been verified by temperature dependent transport
measurements. We calculate the change of the valence band to donor
absorption $\Delta\alpha$ due to the electron spin fluctuations in
the style of Dumke \cite{pr132_1998}
\begin{eqnarray}\label{eqn:deltaalpha}
    \Delta\alpha &= &\frac{64 e^2 \langle |p_{cv}|^2\rangle
    E_v^{1/2}}{\epsilon_0 n c m_0^2 \omega (m_e E_D)^{3/2}}
    \left(\frac{m_{hh}^{3/2}}{[1+(m_{hh}E_v/m_e E_D)]^4}-\right.\nonumber\\
  & &  \left.\frac{m_{lh}^{3/2}}{[1+(m_{lh}E_v/m_e
    E_D)]^4)}\right)\sqrt{N}/V \label{Dumke}
\end{eqnarray}
where $e$ is the electron charge, $\langle |p_{cv}|^2\rangle =
2.1\times10^{-48}~\mathrm{kg^2m^2/s^2}$ the average of the squared
matrix element for transitions between Bloch states, $E_v = (\hbar
\omega + E_D -E_g$), $E_D=0.06$~meV the donor binding energy,
$E_g=1.517$~eV the direct band gap energy, $m_{hh}=0.45~m_0$ and
$m_{lh}=0.082~m_0$ the heavy and light hole masses, respectively,
$n=3.6$ the refractive index, $c$ the velocity of light, $\omega$
the light frequency, $m_0$ the free electron mass and
$m_e=0.067~m_0$ the effective electron mass in the conduction
band. The minus sign results from the optical selection rules
which couple for a given circular polarization heavy hole and
light hole to opposite electron spin states. Applying
Kramers--Kronig relation to $\Delta\alpha$ yields the average
change in refractive index for right-- ($\sigma^+$) and
left--circularly ($\sigma^-$) polarized light and thereby the
average Faraday rotation angle. The change of refractive index
$\Delta n$ at 825~nm due to the thermal fluctuations becomes in
our sample $\Delta n \approx 2.4\times 10^{-8}$ and the resulting
Faraday rotation angle $\Theta_{\rm F} = \approx 7\times
10^{-5}$~rad. The calculated maximum of the noise peak for a spin
relaxation time of 45~ns is $9\times 10^{-17}~{\rm V}^2/{\rm Hz}$.
This calculated noise is for the simplicity of the calculation,
which does not include trionic effects \cite{epjb42_63}, and the
uncertainties in the experimental parameters in good agreement
with the measured $1.5\times 10^{-17}~{\rm V}^2/{\rm Hz}$.

All measurements have been carried out at low temperatures where
electrons are localized and the noise signal is temperature
independent. At higher temperatures the donor bound electrons
become delocalized and the distribution of the electrons in the
conduction band follows a Maxwell--Boltzmann distribution. The
average size of fluctuations $\sigma^2_{N^+ -N^-}$ of the absolute
spin orientation is expected to be for the Maxwell--Boltzmann
distribution - as in the localized case - temperature independent
and equal to N. Nevertheless, the Faraday rotation signal should
become temperature dependent since $\Delta\alpha$ broadens and
moves in energy with increasing temperature. The independence of
$\sigma^2$ on temperature will change to a strong temperature
dependence in samples with higher electron densities where the
electron distribution is a Fermi--Dirac distribution $f(E)$ with
\begin{equation}
    \sigma^2_{N^+ - N^-} = V \int^\infty_{E_{\rm gap}} {\rm DOS}(E)
    f(E)(1-f(E))\, dE,
\end{equation}
where DOS is the density of states. For Fermi--Dirac
distributions,
\begin{equation} \nonumber
\lim_{T\to 0}\sigma_{N^+ - N^-}=0 \qquad\textrm{and}\qquad
\lim_{T\to \infty}\sigma_{N^+ - N^-}=\sqrt{N},
\end{equation}
 with a most pronounced temperature dependence between $T=0$ and
 $k_{\rm B} T \le E_{\rm F}$, where $k_{\rm B}$ is Boltzmann's constant and $E_{\rm F}$ is
 the Fermi energy.

\begin{figure}
\includegraphics[angle=0,width=8cm]{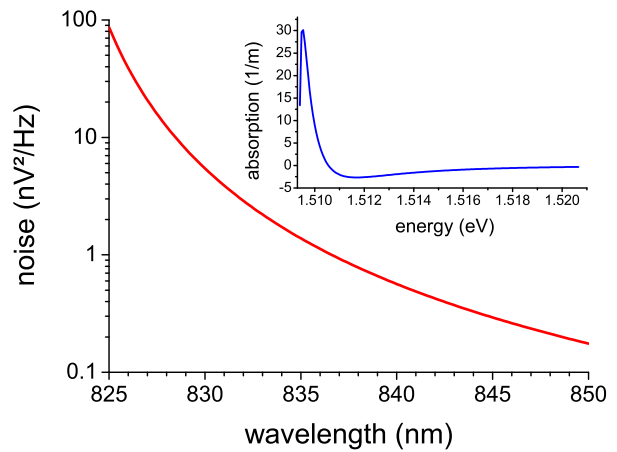}
    \caption{ \label{fig3} (color online) Calculated maximum of the electron spin
    noise signal in
    dependence on detection wavelength for our experimental
    parameters
    The same calculations for $p$--doped GaAs yields a
    signal which is five orders of magnitude smaller due to the
    much faster spin relaxation of free holes.
    The inset depicts the change
    of absorption due to the thermal fluctuations of the electron
    spins in dependence on
    energy [eqn. (\ref{Dumke})].}
\end{figure}

We expect spin noise spectroscopy to have a wide range of
applications in semiconductors. As an example we show in the
following that spin noise is in principle capable to measure
intrinsic electron spin relaxation times at low temperature with
less uncertainties than traditional Faraday rotation or Hanle
measurements. The intrinsic electron spin relaxation time recently
gained new interest, since experiments and calculations by Beck et
al. at 4~K adumbrate that the electron spin relaxation times in
GaAs with an $n$--doping in the range between $10^{15}$~cm$^{-3}$
and $10^{16}$~cm$^{-3}$ might not be limited by the anisotropic
exchange interaction \cite{prb64_075305} but by the
Dyakonov--Perel (DP) mechanism \cite{Beck2005}. The DP mechanism
vanishes for electrons with wave vectors $|\vec{k}|\rightarrow 0$.
Therefore low electron temperatures and the absence of additional
spin relaxation mechanism are required for long spin relaxation
times. These requirements are extremely difficult to achieve in
traditional Faraday rotation and Hanle measurements for two
reasons. Firstly, the carrier temperatures are intrinsically
higher than the sample temperature and secondly, the optically
injected holes cause additional spin relaxation due to the well
known Bir--Aronov--Pikus (BAP) mechanism \cite{BAP}. The
temperature of the optically injected carriers are at low sample
temperatures intrinsically higher than the sample temperature
since even resonant optical excitation (excitation of the
$n$--doped semiconductor at the Fermi edge) yields hot holes with
high $k$ values and carrier cooling at low temperatures is
extremely inefficient due to the inefficient coupling of carriers
with acoustic phonons \cite{prl42_1090}. Secondly, spin relaxation
due to holes by the BAP mechanism is difficult to rule out since
carrier recombination times become extremely long at low carrier
concentration. The importance of BAP even at lowest pump
intensities has been nicely demonstrated by several groups (see
e.g. Ref.~\cite{prb66_245204}, inset of Fig.~2). Spin noise
measurements on the other hand do not necessarily excite carriers
and thereby circumvent the above problems including the sometimes
displeasing dynamical nuclear spin polarization. Since the
amplitude of the spin noise signal is proportional to the spin
relaxation time, spin noise spectroscopy particularly qualifies
for systems with long spin relaxation times.

To evidence that spin noise measurements are sufficiently
sensitive even below the band tail absorption, we have calculated
the Faraday rotation noise signal in dependence on wavelength (see
Figure~\ref{fig3}). The noise power decreases by two order of
magnitude when we increase the excitation wavelength from 825~nm
to 840~nm. At 825~nm about 90~\% of the light is absorbed in our
370~$\mu$m thick GaAs sample yielding an excitation density of
$5\times 10^{14}~{\rm cm}^{-3}$ assuming a radiative carrier life
time of 10~ns. At 840~nm the absorption is nearly negligible. We
have also calculated the Faraday rotation signal for non-localized
electrons, e.g. in modulation doped GaAs, and find at the same
electron temperature a signal decrease of only one order of
magnitude at 840~nm and two orders of magnitude at 900~nm. We have
to compare these decreases in noise power signal with the
signal--to--noise ratio in Fig.~\ref{fig1}. Keeping in mind the
deficiencies of our present experiment we are confident that even
two orders of magnitude lower doping concentrations should be
measurable since the noise signal decreases only with the square
root of the doping concentration.

In conclusion, we have measured  the spin dynamics in $n$--doped
bulk GaAs by spin noise spectroscopy. The sensitive technique
allows the disturbance free measurement of the spin dynamics in
semiconductors with high accuracy. The measurements show in
combination with calculations that spin noise spectroscopy in
semiconductors is a powerful tool, circumvents common problems of
carrier heating and injection of interfering holes, and promises
new insight into spin relaxation, magnetization dynamics, and
electron--spin correlations.

This work has been supported by the Deutsche
Forschungsgemeinschaft, the Alexander--von--Humboldt foundation,
and the German Federal Ministry of Education and Research (BMBF)
by the funding program NanoQUIT. We thank R.~Winkler for helpful
comments. M.O. is indebted to Axel Lorke and Pierre M. Petroff for
stimulating discussions and enjoyable hospitality at the UCSB.


\begin{references}
\bibitem{note_electrical_noise} The spectrum analyzer exhibits a
        strong additional periodic structure on top of the power spectrum
        which is independent of the input signal and probably
        arises from digitalization errors. This additional
        structure is about one order of magnitude stronger than
        the electrical noise but can be also fairly reliably subtracted.
\end{references}
\end{document}